\documentstyle[pre,aps,epsfig,floats,twocolumn]{revtex} 

\begin{document}
\small
\draft

\title{Mean-Field Approximations to the Longest Common Subsequence Problem}

\author{J.~Boutet de Monvel}
\address{ENT Laboratory - King Gustav V Research Institute\\ 
Karolinska Hospital, 17 176 Stockholm\\
Email: Jacques.Boutet.de.Monvel@ood.ki.se}

\maketitle

\begin{abstract}
The Longest Common Subsequence (LCS) problem is a fundamental problem of sequence comparison. A natural 
approximation to this problem is a model in which every pairs of letters of two ``sequences'' are matched 
{\it independently} of the other pairs with probability $1/S$, $S$ representing the size of the alphabet. 
This model is analogous to a mean field  version of the LCS problem, which can be solved with a cavity 
approach \cite{Boutet99_EPJB}. We refine here this approximation by incorporating in a systematic 
way correlations among the matches in the cavity calculation. We  obtain a series of closer and 
closer approximations to the LCS problem, which we quantify in the large $S$ limit, both with a 
perturbative approach and by Monte-Carlo simulations. We find that, as it happens in the expansion 
around mean-field for other disordered systems, the corrections to our approximations depend upon 
long-ranged correlation effects which render the large $S$ expansion non-perturbative.
\end{abstract}

\section{Introduction}
The Longest Common Subsequence (LCS) problem is a simple and fundamental example of a sequence comparison problem. Such problems arise under various important situations, ranging from biology to combinatorics and computational sciences \cite{SankoffKruskal83_Book}. A frequent problem of molecular biology is the detection of evolutionary relationships between different molecules \cite{Waterman89_Book}: Given two DNA molecules which evolved from a common ancestor through a process of random insertions and deletions, how can one recover the ancestor? A possible answer is to solve a particular instance of the LCS problem, namely to look for sequences of nucleotides 
that appear {\it in the same order} in the two DNA molecules, and to pick such a common subsequence that is as long, {\it i.e.} contains as many nucleotides, as possible. Replacing the two DNA molecules by two general sequences $X=(X_1,...,X_N)$ and $Y=(Y_1,...,Y_M)$ (not necessarily of equal lengths) taken from a given alphabet, one obtains a general instance of the LCS problem. As it is natural to expect, when $X$ and $Y$ are very long sequences whose elements are taken at random independently from an alphabet of $S$ letters (with $S\ge 2$), there is a definite density of matched points in a LCS of $X$ and $Y$. More precisely if $L_N$ denotes the length (the number of letters) of a LCS of $(X_1,...,X_N)$ and $(Y_1,...,Y_N)$, one can prove (see {\it e.g.} \cite{Steele97_Book}) that with probability one, $L_N/N$ tend to a non random constant $\gamma_S$ as $N\to \infty$. The determination of $\gamma_S$ and of the rate at which $L_N/N$ approaches this limit are much studied combinatorial problems \cite{ChvatalSankoff75_JAP,DancikPaterson94_STACS,Alexander94_AAP}. 
A connection with statistical physics has been provided by Hwa and L\H{a}ssig \cite{HwaLassig96_PRL} who found that Needleman-Wunsch sequence alignment, a popular comparison scheme for DNA and proteins of which the LCS problem is a special case \cite{NeedlemanWunsch70_JMB}, falls in the universality class of directed polymers in a random medium. This connection is based a geometric interpretation (explained in the next section) of the LCS problem as a longest path problem \cite{Ukkonen85_IC}. The randomness in the above ``Random String'' model can be encoded in variables $\epsilon_{ij}$ defined as occupation numbers for the matches of $X$ and $Y$, namely $\epsilon_{ij}=\delta_{X_i,Y_j}=1$ if $X_i=Y_j$ and $0$ otherwise. The presence of long-ranged correlations among the matches (for example given any indices $i_1,j_1,i_2,j_2$, the variables $\epsilon_{i_1j_1},\epsilon_{i_1j_2},\epsilon_{i_2j_1},\epsilon_{i_2j_2}$ are obviously correlated) complicates the problem very much, and to date the computation of the average length of a LCS has turned out to be intractable. In \cite{Boutet99_EPJB}, we studied a related ``Bernoulli Matching'' model where the $\epsilon_{ij}$'s are taken to be independent and identically distributed random variables with $P(\epsilon_{ij}=1)=1-P(\epsilon_{ij}=0)=1/S$. It turns out that this model is very analogous to a mean field version of the LCS problem, which can be solved using a cavity approach. This solution was found 
to provide a very good approximation (whose precision ameliorates as the size of the alphabet increases) to the average LCS length of two random strings measured from direct Monte Carlo simulations.
We pursue here the work of \cite{Boutet99_EPJB} by studying the behaviour of the above ``mean field'' approximation in the limit of large alphabets. We describe a method which allows to refine the cavity calculation made for the Bernoulli Matching model, by taking correlations of the Random String model into account in a systematic way. This leads to a series of approximations getting closer and closer to the LCS problem, which we quantify within a perturbative approach valid in the limit $S\to \infty$. We find that, while our perturbative approach 
provides an excellent approximation to the LCS problem at finite $S$, it leads to a singular expansion (in powers of $1/\sqrt{S}$) around the Bernoulli Matching model. In particular, the {\it leading} corrections to this mean-field approximation depend upon long-ranged correlation effects among the matches and cannot be captured by the method we use. 

\section{The cavity solution to the Bernoulli Matching model} \label{cavity_solution}
Consider the lattice ${\cal C}_{NM}$ formed by the integer points $(ij)$, 
$0\le i\le N, 0\le j \le M$ together with nearest neighbor bonds, and add a diagonal bond $\{(i-1,j-1),(ij)\}$ for each point $(ij)$ such that $\epsilon_{ij}=1$ (we call such a point a {\it match}). Define the weight of any path on ${\cal C}_{NM}$ to be the number of diagonal bonds that it contains, and let $L_{ij}$ be the maximum possible weight of a directed path joining the point $(0,0)$ to $(ij)$. In the Random String model $L_{ij}$ is just the length of a LCS of the substrings $(X_1,...,X_i)$ and $(Y_1,...,Y_j)$. 
Setting $L_{i,0}=L_{0,j}=0$, the $L_{ij}$'s satisfy the following recursion relation:
\begin{equation} \label{Lij_rec}
L_{ij}=\max(L_{i-1,j},L_{i,j-1},L_{i-1,j-1}+\epsilon_{ij}),
\end{equation}
which follows from the fact that any directed path ending at $(ij)$ must visit one of the points $(i-1,j),(i,j-1)$ or $(i-1,j-1)$. It turns out to be more convenient to work with the {\it local gradient} variables 
$\nu_{ij}=L_{ij}-L_{i-1,j}$ and $\mu_{ij}=L_{ij}-L_{i,j-1}$, rather than with $L_{ij}$ itself. It is obvious from (\ref{Lij_rec}) that $\nu_{ij}$ and $\mu_{ij}$ can take only the values $0$ or $1$. Writing $\bar{x}=1-x$ if $x\in \{0,1\}$, the recursion relations for $\nu_{ij}$ and $\mu_{ij}$ can be written in algebraic form:
\begin{eqnarray}\label{numu_rec}
\nu_{ij}=(1-\bar{\epsilon}_{ij}\bar{\nu}_{i,j-1})\bar{\mu}_{i-1,j} \nonumber\\
\mu_{ij}=(1-\bar{\epsilon}_{ij}\bar{\mu}_{i-1,j})\bar{\nu}_{i,j-1}
\end{eqnarray}
with $\nu_{i,0}=\nu_{0,i}=\mu_{i,0}=\mu_{0,i}=0$.
The key property which was used (but left unjustified) in \cite{Boutet99_EPJB} is that in the Bernoulli Matching model the variables $\nu_{ij}$ and $\mu_{ij}$ along $i+j=t$ become {\it independent} in the limit $t\to \infty$. This can be viewed as a consequence of the directed polymer picture of \cite{HwaLassig96_PRL}, if we interpret $L_{ij}$ as the height profile $L(x,t)$ (as a function of $x=i-j$ and $t=i+j$) of a growing 1D interface, described in a continuum limit by the Kardar-Parisi-Zhang equation (KPZ) \cite{KPZ86_PRL}. In this limit it is known \cite{KrugSpohn91_inbook} that the gradient of $L(x,t)$ become decorrelated along $x$ as $t\to \infty$ \footnote{The author is grateful to R. Bundschuh for pointing this out to him.}. The $\nu_{ij}$'s and $\mu_{ij}$'s could still have finite ranged correlations along the $x$ direction at the {\it discrete} level of the model, however this does not happen here. This can be seen from a Markov chain approach which we present in the appendix. The consequence of this decorrelation property is that we can use eqs. (\ref{numu_rec}) in a self-consistent way in order to compute the probabilities $p_{ij}=P(\nu_{ij}=1)$ and $p'_{ij}=P(\mu_{ij}=1)$ for $i,j$ large. In this sense we may view the Bernoulli Matching model as a mean field model in which (\ref{numu_rec}) are ``cavity equations'' \cite{Boutet99_EPJB}. Assuming independance of $\nu_{i-1,j},\mu_{i,j-1}$ and $\epsilon_{ij}$ in (\ref{numu_rec}) we get
\begin{eqnarray}\label{ppij_rec}
p_{ij}=1-p'_{i-1,j}-(1-1/S)(1-p_{i,j-1})(1-p'_{i-1,j}) \nonumber \\
p'_{ij}=1-p_{i,j-1}-(1-1/S)(1-p_{i,j-1})(1-p'_{i-1,j}).
\end{eqnarray}
These equations can be solved in a continuum limit \cite{Boutet99_EPJB}, 
leading to 
\begin{equation} \label{ppr_eq}
p(r)={\sqrt{rS}-1\over S-1}, \indent p'(r)={\sqrt{S/r}-1\over S-1}
\end{equation}
where $p(r)=\lim_{i\to \infty} p_{i,ri}$ and $p'(r)=\lim_{i\to \infty} p'_{i,ri}$, and
\begin{equation} \label{gammaB}
\gamma^B_S(r)=\lim_{i\to \infty}{L_{i,ri}\over i}=p(r)+rp'(r)={\sqrt{rS}-r-1\over S-1}.
\end{equation}
Note that (\ref{ppr_eq}) and (\ref{gammaB}) are only valid for $1/S\le r\le S$. If $r>S$, resp. $r<1/S$, the process evolves towards the state $(p,p')=(1,0)$, 
resp. $(p,p')=(0,1)$ (this is a ``percolation transition'' of the LCS 
problem \cite{Boutet99_EPJB}). 


\section{Bernoulli Matching model versus Random String model}
\begin{figure}
\begin{center}
\resizebox{0.45\textwidth}{!}{\includegraphics{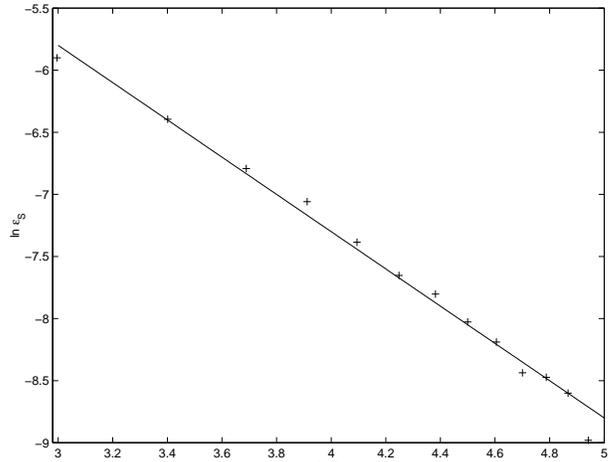}}
\caption{Scaling of $\epsilon_S=\gamma^B_S-\gamma_S$ with $S$. Log-log plot for $20\le S\le 130$ (error bars not reproduced), together with a reference line of slope $-3/2$.} 
\label{fig2}
\end{center}
\end{figure}

Let us brieffly compare eq. (\ref{gammaB}) to the numerical estimates obtained for the Random String model. For simplicity we shall restrict to the case $r=1$ (random strings of equal sizes). Using Monte Carlo simulations and a finite size scaling analysis \cite{Boutet99_EPJB} it was found that the relative error  $(\gamma^B_S-\gamma_S)/\gamma_S$ (with $\gamma^B_S=\gamma^B_S(r=1)=2/(1+\sqrt{S})$) is about $+2\%$ for $S=2$ and $S=3$, and decreases for $4\le S\le 15$ (it is about $+0.9\%$ for $S=15$). 
Figure \ref{fig2} reproduces the behaviour of the difference 
$\epsilon_S=\gamma_S^B-\gamma_S$ in a log-log plot for $S$ up to $130$. Numerically $\epsilon_S=\gamma^B_S-\gamma_S$ decreases rather fast at large $S$, showing a $1/S^{\alpha}$ dependance for a value of $\alpha$ compatible with $3/2$. We remark that a simple expansion holds for the Bernoulli Matching model, as we have $S\gamma^B_S/(2\sqrt{S}-2)=1/(1-1/S)=1+1/S+1/S^2+...$. Anticipating on a similar expansion for the Random String model we would expect corrections in the left-hand-side of this relation to occur in the $1/S$-term.

\section{Incorporation of correlations}
\begin{figure}
\begin{center}
\resizebox{0.45\textwidth}{!}{\includegraphics{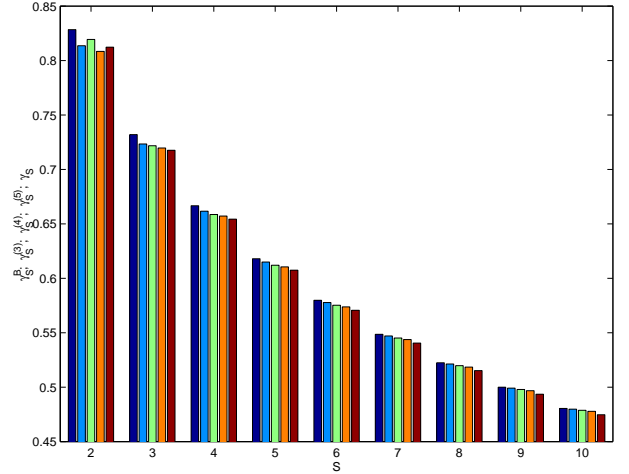}}
\caption{Perturbative approximations to $\gamma_S$. This is a bar graph: For each $2\le S\le 10$, the first to 
fifth bars from left to right give respectively the values of $\gamma^B_S=\gamma^{(1)}_S$, $\gamma^{(3)}_S$, 
$\gamma^{(4)}_S$, $\gamma^{(5)}_S$, and our numerical estimate of $\gamma_S$.} 
\label{fig3}
\end{center}
\end{figure}

We now come to the question of computing corrections to the above approximation, by incorporating some of the 
correlations of the Random String model in our calculation. This can be done in a systematic way as follows. 
We iterate relations (\ref{numu_rec}) a certain number, say $k$ of times. The resulting equations are averaged,
 taking into account correlations among the $\epsilon_{ij}$'s, to built up the transition probabilities of a 
Markov process which we use as a refined approximation to the LCS problem. This approach is similar to the 
n-tree approximations which were used by Cook and Derrida to obtain a $1/d$ expansion for the directed polymer
 problem on finite dimensional lattices \cite{CookDerrida90_JPA}.
We note however that the Bernoulli Matching model is very different from a model of directed polymers on a 
hierarchical lattice, and the word ``tree'' would be somewhat misleading here. In order to analyse the above 
new process we use a perturbative approach, {\it assuming} that the variables $\nu_{ij}$ and $\mu_{ij}$ for 
$i+j=t$ are independent in the stationary distribution, as they are in the Bernoulli Matching model. 
It is then straightforward to obtain a self-consistent equation for $p=\lim_{i\to \infty} p_{ii}$ 
in the form $f^{(k)}_S(p)=p$, where $f^{(k)}_S(p)$ is an $S$-dependent polynomial of degree $2k$ in $p$. 
The positive solution $p_S^{(k)}$ to this
equation provides with a new approximation ${\gamma}^{(k)}_S=2p_S^{(k)}$ to $\gamma_S$. Since there are no 
$3$-term correlation in the Random String model (correlations among the $\epsilon_{ij}$'s occur only for 
configurations forming loops on the square lattice, {\it e.g.} in the four corners of a rectangle), it follows
that no correlation in the disorder occur at level $k=2$, so ${\gamma}_S^{(k)}$ differ from $\gamma_S^B$ 
only for $k\ge 3$. An explicit computation shows that the equation $f^{(k)}_S(p)=p$ has only one positive 
root, at least up to $k=5$. The corresponding values of $\gamma^{(k)}_S$ thus provide sensible perturbative 
approximations to $\gamma_S$, which are reproduced in figure \ref{fig3}. 
Note that the estimates are improving, at least up to $k=5$ for $S\ge 3$. 
The successive values of $\gamma^{(k)}_2$ are not incompatible with a non monotonous approach to $\gamma_2$. 
The relative error $(\gamma_S^{(k)}-\gamma_S)/\gamma_S$ at $k=5$ is of $-0.48 \%$ for $S=2$ and
$+0.28\%$ for $S=3$, a significant improvement compared 
to the error committed with the Bernoulli Matching estimate $\gamma_S^B$. 
This approximation scheme would be perfectly consistent if a decorrelation property occurred at every levels 
$k$. This is in fact {\it not} the case, for example one can show that in the invariant distribution of the 
process at level $k=3$, the variables $\nu_{ij}$ and $\mu_{ij}$ are necessarily correlated. In the KPZ picture
we may say that for $k\ge 3$, there remains as $t\to \infty$ short-ranged correlations along the 
$x$-direction in the local gradients of the growing interface's height. However these correlations turn out to
 be numerically very small, which explains why our perturbative approach gives already a pretty accurate 
result at $S=2,3$. Moreover when $S$ becomes large this approach becomes more and more accurate, as the exact 
invariant distribution ressembles more and more that of the Bernoulli Matching model, and we expect that the 
leading corrections introduced {\it at a given level $k$} are captured by this approximation.
We now evaluate the behaviour of $\gamma_S^{(k)}$ as $S\to \infty$. This evaluation involves comparing 
$f^{(k)}_S(p)$ to the analogous polynomial $f^{B(k)}_S(p)$ computed within the Bernoulli Matching model. The 
coefficients of $\delta f^{(k)}_S=f^{B(k)}_S-f^{(k)}_S$ are directly related to correlations among the 
$\epsilon_{ij}$'s. For example the computation of $\delta f^{(3)}_S$ involves the $4$-correlation term 
$<\bar{\epsilon}_{i_1 j_1}\bar{\epsilon}_{i_1 j_2} \bar{\epsilon}_{i_2 j_1} \bar{\epsilon}_{i_2 j_2}>
= (1+1/(S-1)^3) (1-1/S)^4$, and we have 
\begin{equation}
\delta f^{(3)}_S(p)={1\over (S-1)^3} (1-{1\over S})^4 (1-p)^2 (1-f_S^{(1)}(p))^2
\end{equation}
with $f_S^{(1)}(p)=f_S^{B(1)}(p)=1-p-(1-1/S)(1-p)^2$.
The coefficients of $\delta f_S^{(k)}$ all turn out to be of order $O(1/S^3)$ or smaller. For completeness
we also give the expression of the polynomial $f_S^{B(3)}(p)$, which reads
\begin{eqnarray}\label{f3b}
f_S^{B(3)}(p)= 1-f_S^{(2)}(p) -(1-{1\over S}){1\over S} (1-p)^2\nonumber\\
-2(1-{1\over S})^2 p(1-p)(1-f_S^{(1)}(p))\nonumber\\
-(1-{1\over S})^3 (1-f_S^{(1)}(p))^2\big(p^2 + (1-{1\over S})(1-p)^2\big),
\end{eqnarray}
where $f_S^{(2)}(p)=f_S^{B(2)}(p)=f_S^{(1)}\circ f_S^{(1)}(p)$.
If we now let $\delta p^{(k)}_S=p^{B(k)}_S-p^{(k)}_S$ where $p^{B(k)}_S$ is the positive solution to 
$f^{B(k)}_S(p)=p$, i.e. $p^{B(k)}_S=p^B_S=1/(1+\sqrt{S})$, a standard computation leads to
\begin{equation}\label{delp}
\delta p^{(k)}_S={\delta f^{(k)}_S(p^B_S)\over 1-{d\over dp}f^{B(k)}_S(p^B_S)}
\end{equation}
up to negligible terms at large $S$. It can be checked that 
${d\over dp}f^{B(k)}_S(p^B_S)=1-2k/\sqrt{S}+O(1/S)$. 
It follows then from (\ref{delp}) that for fixed $k$, the correction $\delta p^{(k)}_S$ is of order 
$O(1/S^{5/2})$, which cannot account for the observed $1/S^{3/2}$ behaviour of 
$\epsilon_S=\gamma^B_S-\gamma_S$. The computation gives $\delta p^{(3)}_S\sim 1/6S^{5/2}$, 
$\delta p^{(4)}_S\sim 1/2S^{5/2}$, and $\delta p^{(5)}_S\sim 1/S^{5/2}$, together with correcting terms in the 
form of series in powers of $1/\sqrt{S}$. We could not extract the general terms of these series for arbitrary 
$k$, but we strongly suspect that (at least) the coefficient $A_k$ in front of $1/S^{5/2}$ {\it diverges} at 
large $k$. The argument goes roughly as follows. The correlation terms involved in
the computation of $\delta f_S^{(k)}(p)$ can all be put into the form 
\begin{equation}
<{\bar \epsilon}_{i_1j_1}...{\bar \epsilon}_{i_lj_l}>=
<(1-\epsilon_{i_1j_1})...(1-\epsilon_{i_lj_l})>.
\end{equation} 
Consider any such correlation term, and let us expand the product and take averages, up to order $1/S^3$. 
It is not difficult to see that the result is 
\begin{eqnarray}
<(1-\epsilon_{i_1j_1})...(1-\epsilon_{i_lj_l})>=\nonumber\\
=1-{l\choose 1}{1\over S}+{l\choose 2}{1\over S^2}-
{l\choose 3}{1\over S^3}+{n_{i_1j_1...i_lj_l}\over S^3}+O({1\over S^4})\nonumber\\
=(1-{1\over S})^l(1+{n_{i_1j_1...i_lj_l}\over S^3}+O({1\over S^4})).
\end{eqnarray}
where $n_{i_1j_1...i_lj_l}$ is the number of rectangles that can be formed with 4 corners on the graph 
made up by the lattice points $(i_1j_1),...,(i_lj_l)$. At level $k$ we have to consider rectangles formed on
the triangular lattice $\Delta_k$ made up by the points $(i,j)$ such that $0\le i,j\le k$ and $i+j \ge k$. 
The number $n_k$ of these rectangles satisfies the recursion relation 
$n_k=2n_{k-1}-n_{k-2}+k(k+1)/2$, which in the large $k$ limit gives ${d^2\over dk^2}n_k\sim k^2/2$, 
leading to $n_k\sim {1\over 24}k^4$ (from a more precise computation, taking account of $n_0=0$ and $n_1=1$, 
one finds that $n_k={1\over 24}k^4+{1\over 4}k^3+{11\over 24}k^2+{1\over 4}$). 
All these rectangles are involved in $A_k$, as for example the polynomial $\delta f_S^{(k)}(p)$ 
always contains a term of the form:
\begin{equation}
\big((1-{1\over S})^{k(k+1)\over 2}-<\prod_{(ij)\in \Delta_k}{\bar \epsilon_{ij}}>\big)(1-p)^{2k}
\end{equation}
which gives a contribution $n_k/S^3+O(1/S^{7\over 2})$ to $\delta f_S^{(k)}(p_S)$. 
Unless some special cancellation occurs between the different correlation terms, we thus expect that 
the behaviour of $A_k$ will be approximately given by $A_k\propto n_k/2k$, {\it i.e.} we find
that it diverges like $k^3$ at large $k$.
We conclude that the leading corrections to the Bernoulli Matching model in a large $S$ 
expansion depend on long-ranged correlations among the matches in the Random String model. In order to 
capture them we should look at arbitrarily large values of $k$, for which the above
perturbative approach is no longer valid.  

\section{Conclusion}
The main point of this paper is that, while the Bernoulli Matching model provides a natural and accurate 
mean-field like approximation to the LCS problem valid in the limit of a large alphabet, the corresponding 
large $S$ expansion is non-perturbative: Inclusion of finite-ranged correlations leads to a series with 
diverging coefficients, while the overall behaviour of the expansion at large $S$ does not reproduce the 
observed gap between the two models. This contrasts with the results of \cite{CookDerrida90_JPA}, where the 
n-tree approximation led to a consistent $1/d$ expansion for the directed polymer problem. As already pointed 
out, we are dealing here with a different kind of mean-field approximation. The Bernoulli Matching model is 
not an infinite dimensional model, and replica symmetry is not broken in this model \cite{Boutet99_EPJB}. 
Note that the $1/d$ expansion for the directed polymer problem is also known to be singular, but in a more 
subtle way: Replica symmetry is restored at finite dimensions, leading to important ``tunnelling'' effects 
between the energy valleys of the mean-field picture \cite{ParisiSlanina99_EPJB}. 
An interesting feature of the LCS problem is that the corrections induced by finite-ranged correlations, while
 singular, remain within a perturbative series in powers of $1/\sqrt{S}$. In this respect the situation for 
the LCS problem seems more favourable than in other combinatorial problems where the correlations in the 
disorder induce non-perturbative corrections in an expansion around the mean-field approximation 
\cite{HBM98_EPJB}. This makes the LCS problem an interesting model for investigating this kind of singularity. 

\section*{Appendix.- Markov chain approach to the Bernoulli Matching model}
Let us denote by $(\nu\mu)^t=\{\nu_{ij},\mu_{ij},i+j=t\}$ the {\it state} of the process defined by 
(\ref{numu_rec}), with $t$ interpreted as time. In the Bernoulli Matching model the evolution is Markovian, 
{\it i.e.} the transition from a given state at time $t$ to another state at time $t+1$ is not affected by 
the states at times $t'<t$ (this is not the case in the Random String model, where the transition from time 
$t$ to time $t+1$ is affected by the whole history of the process). 
We will first show that, as a Markov process, the Bernoulli Matching model admits invariant distributions in
which the components of $(\nu\mu)^t$ are completely decorrelated (we mention that the same result has been 
found in a different way in \cite{BundschuhHwa99_inproc}, in the case $N=M$). Consider the relations 
(\ref{numu_rec}) restricted to a given cell of the lattice ${\cal C}_{NM}$. The corresponding ``one-cell''
transition probability $P_1(\nu,\mu|\nu',\mu')$ is given by
\begin{eqnarray}
P_1(\nu,\mu|\nu',\mu')= {\nu \mu \bar{\nu}' \bar{\mu}'\over S} + 
\bar{\nu} \mu \bar{\nu}' \mu' + \nu \bar{\mu} \nu' \bar{\mu}' \nonumber \\
+\bar{\nu}\bar{\mu}(\nu'\mu'+(1-1/S)\bar{\nu}'\bar{\mu}').
\end{eqnarray}
A simple computation shows that the one-cell Perron-Frobenius equation
$P_1.\pi_1=\pi_1$ (with matrix notations) has a solution of the form
$\pi_1(\nu,\mu)=(p\nu+(1-p)\bar{\nu})(p'\mu+(1-p')\bar{\mu})$
provided the probabilities $p$ and $p'$ satisfy 
\begin{equation}\label{pp_eq}
1=p+p'+(S-1)pp'. 
\end{equation}
Suppose now that we let the bonds on the lower corner of any given rectangle be occupied independently with 
probability $p$ for horizontal bonds and $p'$ for vertical bonds. A moment's thought shows that the {\it same} 
distribution will hold for the upper corner bonds if we let the occupation numbers for the bonds inside the 
rectangle evolve according to (\ref{numu_rec}), as long as $p$ and $p'$ satisfy (\ref{pp_eq}). Hence any 
solution of (\ref{pp_eq}) provides with a
decorrelated invariant distribution as was claimed. In a continuum limit, these invariant distributions 
can be identified locally with the ``pure'' invariant distributions of the process, i.e. those 
invariant distributions evolved from a single initial state of the variables $\nu,\mu$. 
More precisely let us impose periodic boundary conditions along the $x=i-j$ direction 
(this is a way of working ``locally''). We let $\nu^t=(\nu_1^t,...,\nu_L^t)$ and $\mu^t=(\mu_1^t,...,\mu^t_L)$ 
be the state variables at time $t$ on a band of ``width'' $L$, and we adopt a numerotation such that 
(\ref{numu_rec}) reads 
\begin{eqnarray} \label{numu_rec_L}
\nu_i^{t+1}=(1-\bar{\epsilon}_i^{t+1}\bar{\nu}_{i+1}^t)\bar{\mu}_i^t \nonumber\\
\mu_i^{t+1}=(1-\bar{\epsilon}_i^{t+1}\bar{\mu}_i^t)\bar{\nu}_{i+1}^t,
\end{eqnarray}
for $i=1,...,L$ ($L+1$ being identified with $1$).
From the remarks made above the Perron-Frobenius equation $P_L.\pi_L=\pi_L$ of this process admits solutions 
of the form $\pi_L^{(p,p')}(\nu_1,\mu_1....,\nu_L,\mu_L)= \prod_{i=1}^L \pi_1(\nu_i,\mu_i)$
where again $(p,p')$ is any solution of (\ref{pp_eq}). 
For finite $L$ however  $\pi_L^{(p,p')}$ is not pure. To get an understanding of the pure distributions we 
adopt a lattice gaz point of view, remarking that the quantity 
\begin{equation}
C=\sum_{i=1}^L \nu_i-\mu_i
\end{equation}
is a conserved charge of the evolution (this conservation law is exact only under the above periodic 
boundary conditions). It can also be checked that the Markov process defined by (\ref{numu_rec_L}) connects 
any two states $(\nu_i\mu_i),(\nu_i'\mu_i')$ having the same charge $C$. It follows that there are exactly 
$2L+1$ pure distributions, in correspondence with the possible values $-L\le C\le L$. 
We can extract a formal expression for the pure invariant distribution $\pi_C(\nu_i,\mu_i)$ evolved from an 
arbitrary state of charge $C$, from the ``mixed'' invariant distribution $\pi_L^{(p,p')}(\nu_i,\mu_i)$. 
Namely we have 
\begin{equation} \label{piC}
\pi_C(\nu_i,\mu_i)={\pi_L^{(p,p')}(\nu_i,\mu_i)\delta(C-\sum_i \nu_i+\sum_i\mu_i))\over
Z(C)}
\end{equation}
where $Z(C)$ is a normalization factor. Note that, contrary to the appearances, the right-hand-side
of (\ref{piC}) does {\it not} depend on $(p,p')$ (this can be seen directly from the expression
of $\pi_L^{(p,p')}$ by making use of (\ref{pp_eq})). 
In the limit $L\to \infty$, the fluctuations of $C$ about its mean value $L(p'-p)$ with respect to 
$\pi_L^{(p,p')}$ become negligible, and we expect that the differences between the pure distributions 
$\pi_C$ (more precisely the differences between their finite correlation functions) for which $C/L$ is 
close to $p'-p$ will become unsignificant. This can be checked directly from (\ref{piC}) using a saddle-point 
evaluation. Hence the pure distributions on a periodic band of infinite width can be identified with a continuum 
of decorrelated distributions $\pi^{(p,p')}$, parametrized by the solutions of (\ref{pp_eq}) for which 
$0\le p,p'\le 1$. Returning to the original lattice ${\cal C}_{NM}$, one must take care of the boundary 
conditions imposed along the axes $i=0$ and $j=0$. The process will now develop only locally ({\it i.e.} along
any given direction) according to a distribution of the form $\pi^{(p,p')}$, with $p$ and $p'$ being functions 
of $r=i/j$. The cavity approach of section \ref{cavity_solution} allows to treat different boundary conditions in
a simple way. For example, if the horizontal bonds along $j=0$ (resp. the vertical bonds along $i=0$) are 
supposed to be occupied independently with probability $p_1$ (resp. $p_2$), where 
$0\le p_1,p_2\le 1/(1+\sqrt{S})$ (the original problem corresponds to the case $p_1=p_2=0$), one finds that
\begin{eqnarray}
p(r)=p_1, \indent 0\le r\le r_1, \nonumber \\ 
p(r)={\sqrt{rS}-1\over S-1}, \indent r_1\le r\le r_2, \nonumber \\
p(r)={1-p_2\over 1+(S-1)p_2}, \indent r\ge r_2,
\end{eqnarray}
where $r_1$ and $r_2$ are such that  
$p_1=(\sqrt{r_1S}-1)/(S-1), p_2=(\sqrt{S/r_2}-1)/(S-1)$, and $p'(r)$ is such that (\ref{pp_eq}) is satisfied.

\section*{Aknowledgements}
The main part of this work was done at the BiBoS center of Bielefeld. J. BdM. is grateful to professor P. 
Blanchard for his kindness and for several discussions which were very stimulating in this research, and 
he thanks O.C. Martin for many useful comments and suggestions on the manuscript. This work was supported 
by the EU-TMR project ``Stochastic Analysis and its Applications''.

\small
\bibliography{co,jbdm,sgds,seqal}
\bibliographystyle{perrot}
\end{document}